# Lithiation Analysis of Metal Components for Li-Ion Battery using Ion Beams

*Arturo Galindo[a], Neubi Xavier[b], Noelia Maldonado[c], Jesús Díaz-Sánchez[d,e], Carmen Morant[e,f], Gastón García[c], Celia Polop[d,e,g], Qiong Cai[b] and Enrique Vasco[a*]*

a.  *Instituto de Ciencia de Materiales de Madrid, Consejo Superior de Investigaciones Científicas, Madrid, Spain*
b.  *School of Chemistry and Chemical Engineering, Faculty of Engineering and Physical Sciences, University of Surrey, Guildford, United Kingdom*
c.  *Centro de Micro-Análisis de Materiales (CMAM), Universidad Autónoma de Madrid, Madrid, Spain*
d.  *Departamento de Física de la Materia Condensada, Universidad Autónoma de Madrid, Madrid, Spain*
e.  *Instituto Universitario de Ciencia de Materiales Nicolás Cabrera (INC), Universidad Autónoma de Madrid, Madrid, Spain*
f.  *Departamento de Física Aplicada, Universidad Autónoma de Madrid, Madrid, Spain*
g.  *Condensed Matter Physics Center (IFIMAC), Universidad Autónoma de Madrid, Madrid, Spain*

Dated: August 28, 2025

**Corresponding author:**

Enrique Vasco

Instituto de Ciencia de Materiales de Madrid

Consejo Superior de Investigaciones Científicas

CL. Sor Juana Inés de la Cruz 3, Cantoblanco

28049 Madrid, Spain

phone: +34-913348981 / email: enrique.vasco@csic.es

**ABSTRACT**

Metal components are extensively used as current collectors, anodes, and interlayers in lithium-ion batteries. Integrating these functions into one component enhances the cell's energy density and simplifies its design. However, this multifunctional component must meet stringent requirements, including high and reversible Li storage capacity, rapid lithiation/delithiation kinetics, mechanical stability, and safety. Six single-atom metals (Mg, Zn, Al, Ag, Sn and Cu) are screened for lithiation behavior through their interaction with ion beams in electrochemically tested samples subjected to both weak and strong lithiation regimes. These different lithiation regimes allowed us to differentiate between the thermodynamics and kinetic aspects of the lithiation process. Three types of ions are used to determine Li depth profile: $H^+$ for nuclear reaction analysis (NRA), $He^+$ for Rutherford backscattering (RBS), and $Ga^+$ for focused ion beam (FIB) milling. The study reveals three lithiation behaviors: (i) Zn, Al, Sn form pure alloys with Li; (ii) Mg, Ag create intercalation solid solutions; (iii) Cu acts as a lithiation barrier. NRA and RBS offer direct and quantitative data, providing a more comprehensive understanding of the lithiation process in LIB components. These findings fit well with our ab-initio simulation results, establishing a direct correlation between electrochemical features and fundamental thermodynamic parameters.

**Keywords:**

multifunctional metal components for battery

electrochemical metallurgy

ion beam analysis of lithiation

energy materials

# Table of Contents

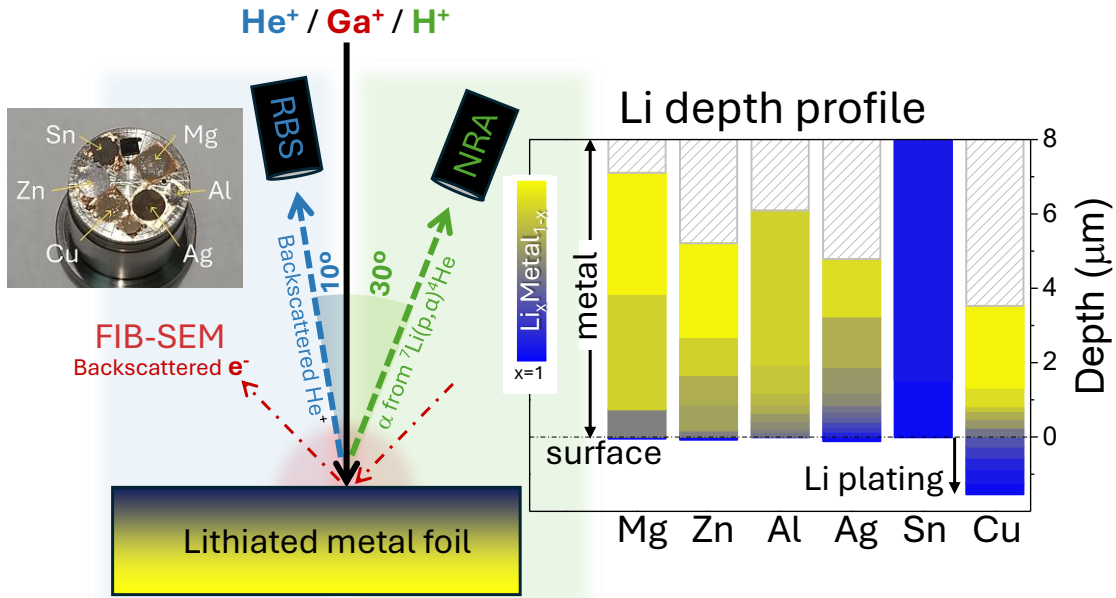

**Figure Caption:** Screening of six electrochemically lithiated metals (Mg, Zn, Al, Ag, Sn, and Cu) utilized as multifunctional components for Li-ion batteries was conducted by ion-beam analytical techniques (nuclear reaction analysis, Rutherford backscattering, and focused ion beam milling-assisted scanning electron microscopy). The study examines the mechanisms of metal lithiation distinguishing between thermodynamics and kinetics, Li depth profiles, and local Li storage capacity.

# 1. Introduction

Metal components have been utilized as anodes in Li-ion batteries (LIBs) due to their low Li oxidation potential.[1–14] They also serve as current collectors that connect the battery to the external circuit[15] allowing the flow of electrons to balance internal ion transport, as well as interlayers or seed-layers to ensure the stability of the electrolyte/anode interface.[16–19] Proposals to integrate all these functionalities into a single metal component would significantly enhance the energy density of the cell and streamline its design.[20,21] During the lithiation of components, mainly those storing Li by conversion and alloying, an early Li accumulation (Li plating) at the interfaces with the electrolyte takes place. Primarily, this issue arises from kinetic limitations in Li intercalation, particularly when the electrochemical current exceeds the rates of Li diffusion or the propagation of the lithiation front into the component bulk.[22–25] Li plating offers both potential advantages and current disadvantages. A significant concern is its susceptibility to the early formation of dendrite, which arises from Li's tendency to resist self-wetting when it grows following a diffusion-limited aggregation (DLA) mechanism by electrodeposition.[26,27] On the other hand, utilizing Li plating to create in-situ anodes for Li-efficient batteries—serving as an alternative to conventional anodes—represents a revolutionary approach in the design of zero Li-excess (also known as anodeless) batteries.[20,28,29] Tailoring the plating process to facilitate via wetting the formation of a uniform metal Li anode—one that exhibits capacity and reversibility comparable to Li-metal batteries—while simultaneously creating a morphologically stable interface with the current collector is a noteworthy challenge.[30]

Before discussing our results, we must define the various lithiation behaviors that were identified in the experiments (see schemes in **Figure 1**). This clarification will help establish consistent terminology throughout our discussion, bridging the gap between the terminologies used in electrochemistry and metallurgy. While electrochemistry refers to the mechanisms of Li storage (intercalation, alloying, plating...), metallurgy defines the relationship between phases within a system under equilibrium conditions (solid solution, intermetallic compound, immiscibility...). When Li accumulates at the interface between the metal component and the electrolyte, it can either diffuse into the bulk of the component or remain at the interface. Its behavior is governed by the balance between adsorption and absorption energies, which is also influenced by the electrochemical potential of the cell (notably, the plating occurs via an electrodeposition process) and the gradient of Li concentration within the component.

In cases where absorption becomes the favored mechanism, the early diffusion of Li into the component bulk leads to the formation of **solid solutions $Li_xM_{1-x}$**, such as substitutional or interstitial Li within the metal host lattice, depending on the differences in their Goldschmidt atomic diameters. The formation of dilute solid solutions (or intercalation compounds, in electrochemical terms) does not produce structural changes or significant volumetric expansions within the component bulk. Additionally, the difference in electronegativity between the metals forming the solid solutions induces intrinsic strains that accumulate as Li fraction (*x*) increases.[31–33] Once the strain energy surpasses the barrier needed for restructuring the component bulk, recrystallization occurs, leading to the formation of **alloys** (or intermetallic compounds, in metallurgical terms). In this context, alloys are defined as phases composed of the two metals (i.e., $Li - M$) arranged into distinctive crystal structures, which are different from those of each individual metal. In general, alloys have larger unit cell volumes than solid solutions, which are governed by the miscibility of one of the metals (mainly Li) in the host lattice of the other. From an electrochemical perspective, solid solutions result in gradual changes in redox pair potentials (i.e. transient profiles) as Li concentration varies, unlike the alloys, which display distinct and well-defined potentials (i.e., plateau profiles). Thus, solid solutions within alloyed lattices, inducing successive transitions between alloys, cause the electrochemical potential profile to pass through a series of plateaus separated by potential transients.[13] Note that we adopted here the term "alloy" as defined in electrochemistry because it is more restrictive than that used in metallurgy (a mixture of several components with at least one metal).

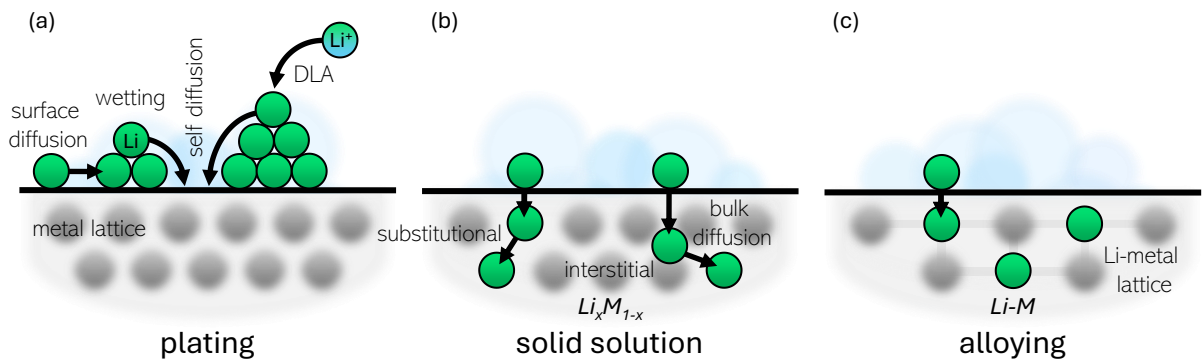

**Fig. 1** Fundamental lithiation mechanisms in metal components: (a) Li plating whose morphology is ruled by the balance between wetting and DLA-induced 3D growth; (b) formation of substitutional or interstitial solid solutions $Li_xM_{1-x}$ (where x denotes the probability of finding Li at the metal lattice sites or the normalized atomic concentration of Li, respectively); and (c) formation of alloys $Li - M$ with Li and metal being arranged into distinctive crystal structures.

In the case of favorable adsorption, the ability of Li plating to wet the metal component surface relies on the interplay of interface energies (specifically those among the electrolyte, the plating, and the component itself) as well as the self-diffusion coefficient of Li. Experimental data indicate that the electrochemical potential drives Li ions to attach to sharper features on rough plating surface.[34,35] Additionally, evidence pointing to a low Li self-diffusion coefficient (~$10^{-11}$ cm$^2$ s$^{-1}$ [25,36]) highlights the challenge of achieving effective wetting during Li plating. This challenge is currently being addressed through innovative surface[37–40] and stress engineering[41,42] approaches. From an electrochemical viewpoint, a plated Li layer exhibiting a roughness greater than the average size of its lateral discontinuities—a necessary morphological condition—will display a lithiation voltage close to zero when referenced to a metal Li electrode. In contrast, the reduction of Li ions on the bare or alloyed surface of the metal component needs higher lithiation voltages.[1] The above-mentioned morphological condition is met for both **continuous plating via wetting** and **rough 3D plating** with dendrite growth. It is important to note that the behavior of Li plating changes as Li concentration increases. It shifts from being influenced by diluted Li-metal bulk interactions to the distinctive growth mode associated with Li bulk. Thus, for example, bulk diffusion of Li driven by concentration depth gradients requires a preliminary adsorbed layer with a threshold thickness that serves as a source of atomic Li for subsequent absorption. Ultimately, beyond the thermodynamic equilibrium between adsorption and absorption energies, which acts as a driving force for lithiation, the bulk diffusion coefficient of Li and/or the propagation rate of the lithiation front within the component volume rule the kinetics of the process.

## 2. Results and discussion

In this study, we investigate the electrochemical lithiation behavior of metal components using a combination of ion beam-based techniques: Nuclear Reaction Analysis (NRA), Rutherford Backscattering Spectroscopy (RBS), and focused ion beam (FIB) milling. Several metals ($M$= Mg, Zn, Al, Ag, Sn, and Cu) were selected based on their distinct lithiation behaviors reported in the literature.[1,2] These techniques provide a different quantitative perspective compared to previous reports,[1,2] where the lithiation habits of selected metals were inferred from their electrochemical behaviors and scanning electron microscopy (SEM) images obtained from backscattered electrons (BSE) in cross-sections uncovered by FIB milling. The information inferred from these standard methods has significant limitations as discussed below. Electrochemical testing of metals vs. Li includes parasitic and other non-faradaic reactions, while FIB milling of the component introduces morphological artifacts to the BSE signal, making it challenging to correlate BSE-SEM contrast with Li gradients. In contrast, we use NRA to directly measure the depth profile of Li concentration via the nuclear reaction $^7$Li(p,α)$^4$He, and complement this with RBS, which offers insights into how Li concentration dilutes the atomic density of the metal components. A detailed description of the NRA and RBS data correlation can be found in the Supporting Information (SI). Although FIB milling with Ga$^+$ is primarily a nanostructuring technique rather than an analytical one, this provides information on the Li depth profile based on the balance between the sputtering yield of the metal component and the Ga$^+$ implantation range. This approach enabled us to qualitatively image the quantitative results of NRA and RBS. The combination of these techniques offers more reliable and comprehensive information, enhancing our understanding of the lithiation phenomenon in metal components. Notably, there are few studies[43–45] to date that employ functional combinations of ion-beam-based techniques to address lithiation mechanisms in LIB components, and even fewer (none to the best of our knowledge) that use these techniques to comparatively study the lithiation of a wide range of metal components in the battery context.

**Figure 2** shows the electrochemical behaviors of selected metals. The evolution of the cyclic voltammetry (CV) currents and galvanostatic charge-discharge (GCD) voltages, used for weak and strong lithiation of the metal components, are illustrated in the left and right columns, respectively. The curves transition from black to red as the number of electrochemical cycles increases. Three distinct types of features can be observed in them: (1) those with diminishing intensities as the cycles progress or exhibiting erratic behavior (marked with asterisks "*" in the CV curves of Mg, Zn, Ag, Sn and Cu, as well as in the GCD curves of Mg); (2) features labeled in green, with increasing peak and plateau shapes in the CV and GCD curves, respectively; and (3) decreasing pseudo-plateaus profiles, labelled in pink, in the GCD curves of Mg and Ag. The first features (1− "*") are attributed to parasitic reactions involving electrolyte, which probably decay as a solid electrolyte interface (SEI) layer is formed. In contrast, the increasing peaks (2) are indicative of the formation of alloys, while the decreasing pseudo-plateaus (3) correspond to solid solutions.[13]

The GCD curves also show, labelled in blue, the electrochemical potentials at which lithiation ($V_l$) and delithiation ($V_d$) occur. When the potentials approach zero (e.g. for Cu), this indicates Li-on-Li processes, whereas voltage plateaus at values greater than zero (e.g., for Al) suggest that the electrochemical reactions take place on the bare, Li-$M$ mixed or alloyed surface of the metal component. The energy difference between lithiation states is defined as the difference in electrochemical potentials multiplied by the charge involved in one mole of the redox reaction ($nF$), where $F$ is the Faraday constant and $n$ is the number of electrons transferred ($n = 1$ for Li). Since alloy states minimize the strain energy associated with Li-rich solid solutions, the delithiation potential of alloys are presumably higher than those of solid solutions. The latter are also expected to change progressively with the chemical environment. These insights allow us to distinguish between alloys (labelled in green) and solid solutions (pink) formed during the electrochemical tests shown in Figure 2, aligning with the results and binary phase diagrams discussed below.

Another important aspect to consider is the kinetics of lithiation, as showcased in the encircled regions in the Al and Ag GCD curves. While lithiation leads to rapid Li accumulation on the Ag surface (with $V_l$ dropping for capacities below 1 mA·h), Li accumulation on the Al surface occurs at significantly higher capacities (>3 mA·h). This observation implies that the rate of propagation of the lithiation front in Al is substantially greater than in Ag, resulting in premature Li saturation in the latter.

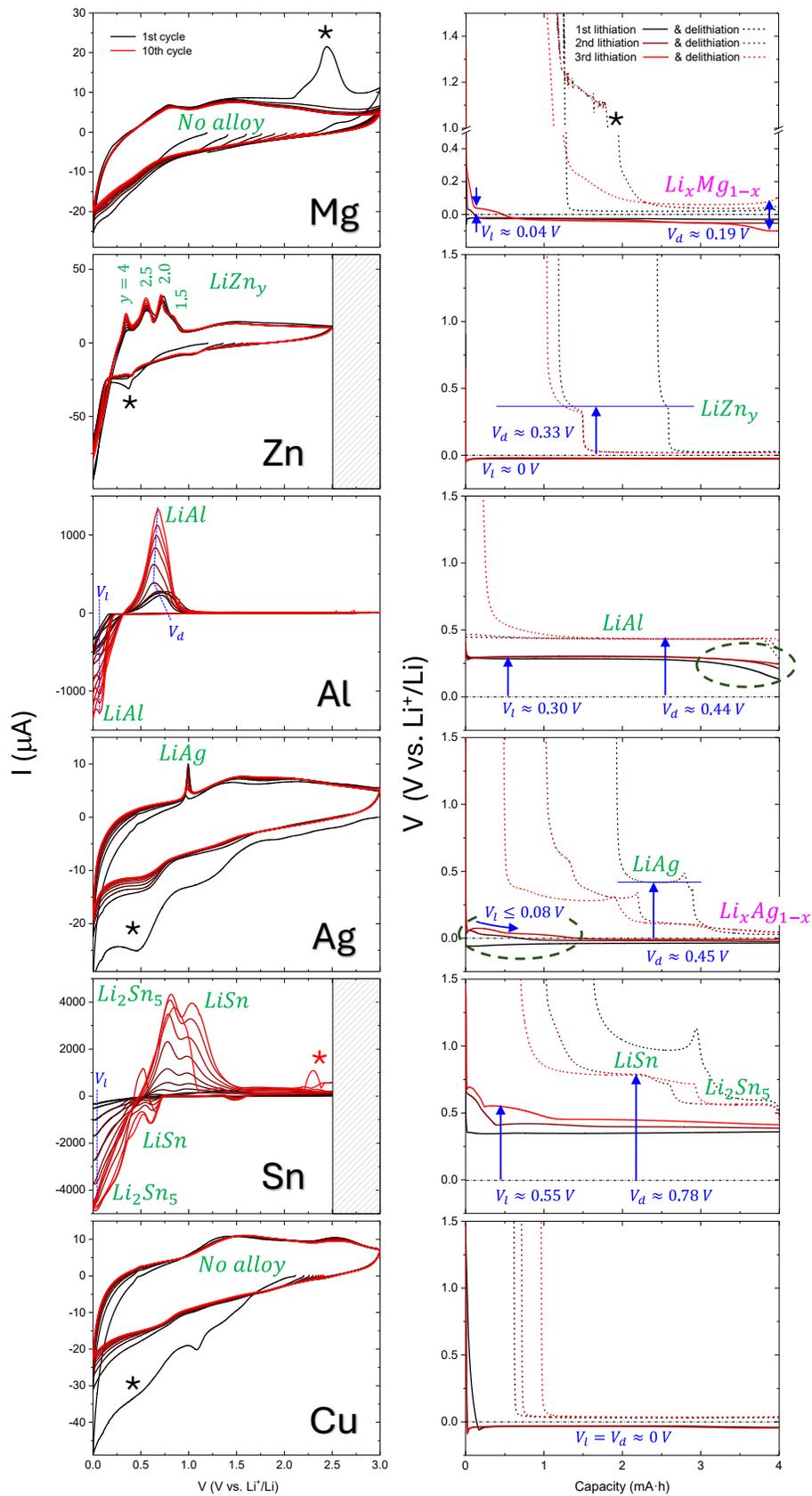

The results presented in Figure 2 align with previous studies on metal component screening for LIBs[1,2] and lead to the following preliminary conclusions: (i) Zn, Al, Ag, and Sn form bulk alloys with Li, whereas Mg and Cu do not. (ii) Li alloys with Zn and Ag undergo an initial stage of Li adsorption (with $V_l \sim 0$), in contrast to the behavior observed for Al and Sn. This behavior is probably due to kinetic limitations in the lithiation process, since the delithiation in Zn and Ag occurs at higher $V_d > 0$. (iii) Mg lithiation occurs following a decreasing pseudo-plateau of $V_l$, which reaches negative values (with respect to the redox potential of metal Li) for high Li contents. $V_l < 0$ rules out surface accumulation of Li. Consequently, since Mg does not form alloy with Li (no peaks in the CV curves), this state likely corresponds to a $Li_xMg_{1-x}$ solid solution. (iv) Li remains adsorbed on the surfaces of Cu with $V_l \approx V_d \approx 0$.

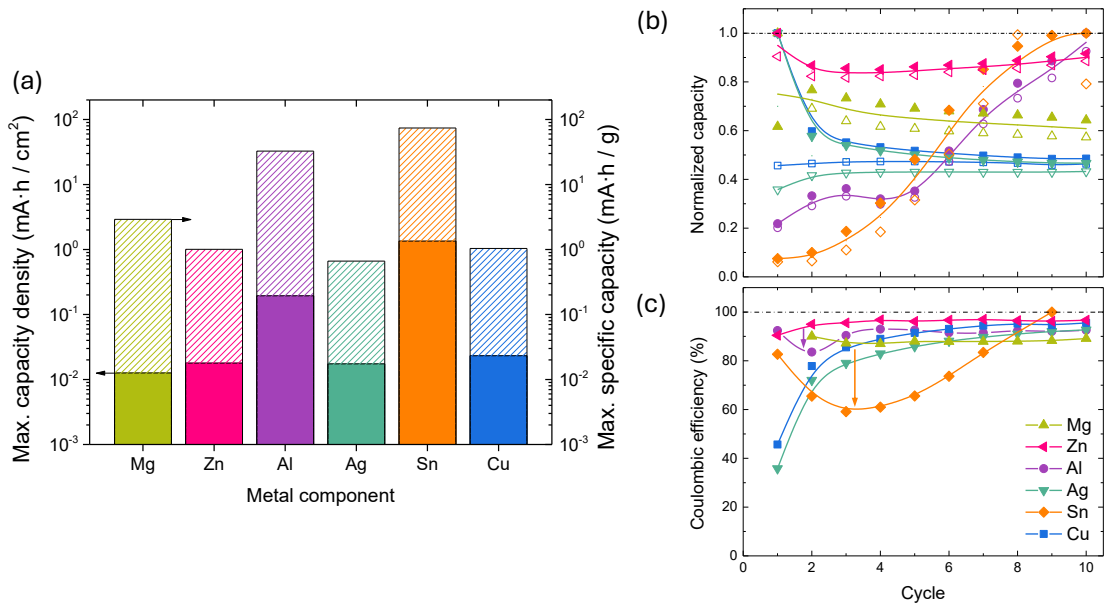

**Fig. 3** Electrochemical magnitudes concerning weak lithiation of metal components: **(a)** maximum capacity density (solid bars—left axis) and its equivalent specific capacity (dashed bars—right axis) considering the total thickness of the components; **(b)** cyclic dependence of the lithiation (solid symbols) and delithiation (open symbols) capacities normalized by the maximum capacity density in (a); and **(c)** cyclic dependence of the Coulombic efficiency (i.e., the ratio of the delithiation capacity to the lithiation capacity). Coulombic efficiencies of cycles dominated by parasitic reactions (e.g. the first cycle in the case of Mg) are omitted.

**Figure 3** illustrates the capacity of metal components in the weak lithiation regime (Figure 3a), emphasizing their normalized evolution (Figure 3b) and Coulombic efficiencies (Figure 3c) over multiple electrochemical cycles. Three distinct behaviors regarding capacity and electrochemical reversibility are observed: (i) Metals with high Li-storage capacities (Al and Sn) show a decrease in their Coulombic efficiencies during the initial cycles (cycles 2-3 for Al and cycles 2-7 for Sn, as indicated by the down arrows in Figure 3c), followed by a recovery. This decrease suggests some degree of irreversibility in the lithiation-delithiation process, likely due to the early formation of alloys with Li, as revealed in Figure 2 for Al and Sn. The simultaneous formation of several alloys, particularly in the case of Sn (e.g., $Li_2Sn_5$ and $LiSn$ in Figure 2), leads to a more significant drop in efficiency, as evidenced by the varying arrow sizes shown in Figure 3c. Component capacity increases with the number of cycles as the alloys stabilize and propagate in their volumes. (ii) Metals with moderate Li-storage capacities (Ag and Cu) show a low Coulombic efficiency during the first cycles, which increases rapidly. This behavior points to an initial adsorption of Li, which is then either retained on the surface (Cu) or absorbed (Ag). Coulombic efficiency recovers around 95% once the SEI is formed, stabilizing the capacity at average values. (iii) Metals with moderate to low Li-storage capacities (Zn and Mg) maintain stable efficiencies around 90%. This stability is observed because the CV curves for Zn and Mg (as shown in Figure 2) exhibit minimal variation once parasitic reactions cease.

In summary, Figure 3 identifies two irreversible processes governing the lithiation of metal components that impact the Coulombic efficiency of these systems: the formation of alloys and the SEI. While the former contributes to enhancing the ultimate battery capacity, the latter provides protection against parasitic reactions between the electrolyte and other components. Notably, metals that quickly form pure alloys with Li (such as Al and Sn) show no evidence of SEI formation. Since electrochemistry does not provide information on the Li depth profile, it indicates that the calculated specific capacities (right axis in Figure 3a) are likely underestimated based on considering a uniform distribution of Li within the component volume. This may also apply to some of the specific capacities reported by Heligman *et al.,*[1] where this effect has been considered to estimate a probable range of component volume utilization. In this work, we directly measure this lithiation range.

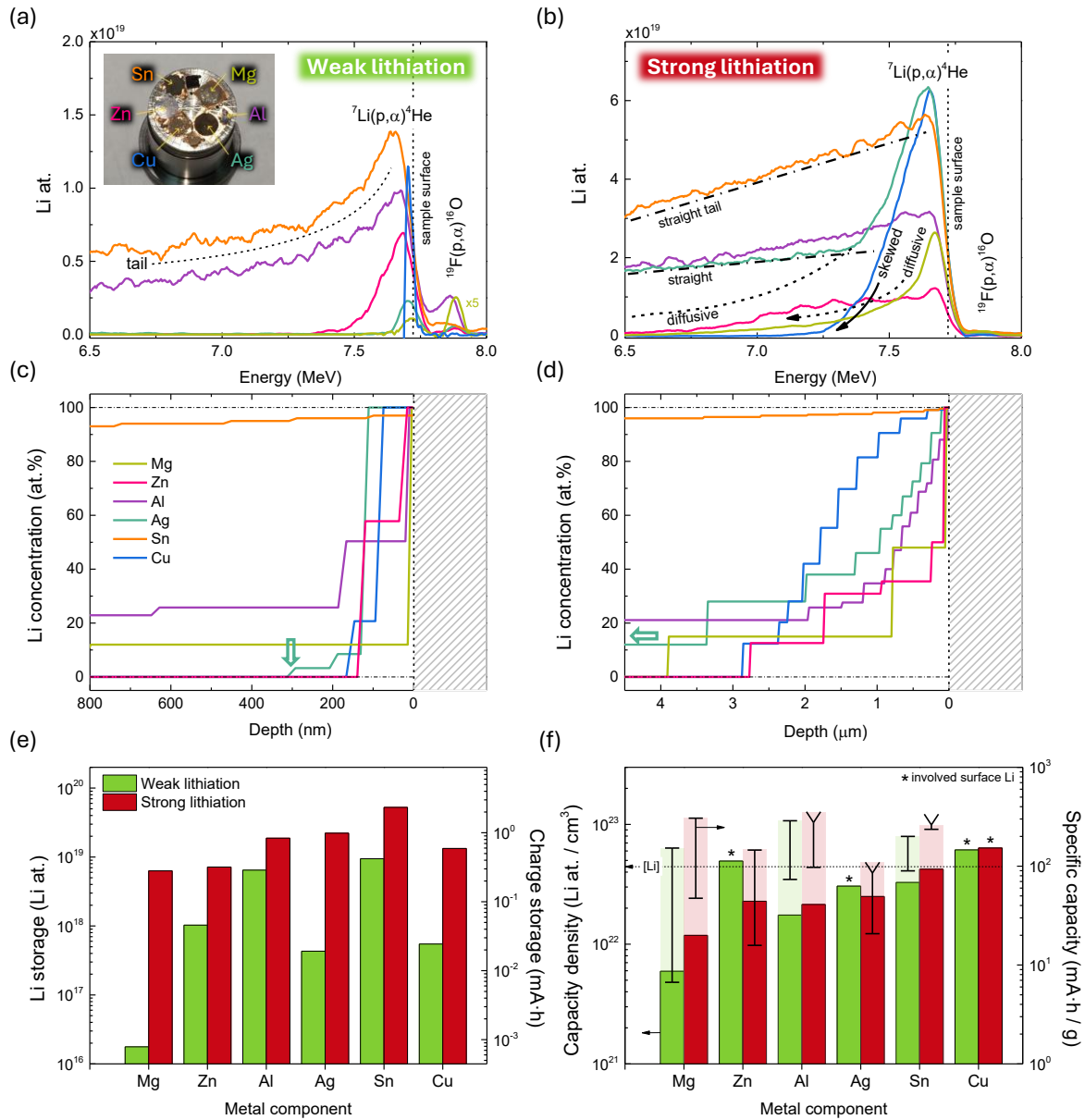

**Fig. 4 (a)** and **(b)** Li NRA bands with 3-MeV H⁺ for weak and strong lithiation regimes, respectively. The signals at higher energies correspond to α particles originating from the nuclear reaction $^{19}F(p,\alpha)^{16}O$, involving fluorine from electrolyte residues. Dashed curves and arrows illustrate the types of tails of the Li NRA bands at lower energies (i.e., the deepest Li layers). The inset in (a) displays lithiated metal components in an NRA-RBS sample holder. **(c)** and **(d)** Li depth profiles obtained from (a) and (b) by fitting the Li NRA bands using the SIMNRA 7.03 code[46,47] and combining them with RBS measurements using He⁺. **(e)** Lithium storage (along with its equivalent charge), corresponding to the areas under the Li NRA bands in (a) and (b). **(f)** Capacity densities (bars referred to the left axis) and specific capacities (vertical line segments, right axis) of metal components calculated from the lithiation ranges of the Li depth profiles in (c) and (d) —upper limit— and the full component volumes —lower limit.

**Figure 4** shows the NRA results for weak (Figure 4a) and strong (Figure 4b) lithiation regimes. The Li NRA bands were analyzed using the SIMNRA 7.03 code[46,47] and combined with RBS measurements (discussed below) to obtain the corresponding Li depth profiles, as displayed in Figure 4c and 4d. The average depth resolutions were estimated to be 270 nm (1260 x $10^{15}$ at. cm$^{-2}$) and 60 nm (325 x $10^{15}$ at. cm$^{-2}$) for our NRA and RBS measurements, respectively, by RESOLNRA 1.11 code.[48]

A qualitative assessment of the Li NRA bands in Figure 4a and 4b reveals different curve shapes, particularly the distinctive tails associated with each lithiation behavior. Under weak lithiation conditions, both Sn and Al display wide tails at low energy (<7.5 MeV), which blend with broad surface peaks (Figure 4a). As Li concentration increases in the strong lithiation regime, these low-energy tails intensify and evolve into straight profiles that reach the metal surface (Figure 4b). Given that Figure 2 indicates Al and Sn are early alloyed with Li, we can attribute the straight tails to a lithiation behavior driven by alloy formation. In this context, Figure 4b shows that the Li NRA band in Zn exhibits a similar shape to those of Sn and Al, albeit with lower intensity. This suggests that the Zn lithiation is also governed by alloy formation, requiring higher Li injections to compensate for its slower diffusion kinetics, a phenomenon also observed by Wang *et al*.[49] On the other hand, Ag and Mg display small Gaussian-shaped peaks at low lithiation (Figure 4a), which increase in intensity and develop low-energy tails (diffusive + straight profile for Ag, and purely diffusive for Mg) as the Li concentration rises (Figure 4b). This diffusive segment of the tails can be attributed to solid solutions (as supported by findings in Figure 2), which are formed via Li diffusion driven by concentration gradients. The transition from a solid solution to an alloy with Li for Ag is determined by the solubility limit, as discussed below. Finally, the narrow surface peak of the Li NRA band in Cu at weak lithiation expands into a broad skewed Gaussian peak at strong lithiation. This change can be attributed to inhomogeneous Li adsorption on the copper surface, resulting in a rough 3D plating. This plating is likely the result of a DLA mechanism operating during electrodeposition. A detailed analysis of the low-energy tails of the Li depth profiles, obtained through complementary measurements of RBS, is presented in the SI.

The quantities of Li stored in each metal component, corresponding to the areas under the Li NRA bands in Figure 4a and 4b, are depicted in Figure 4e along with their equivalent electrical charges (left axis). These charges are consistent with those involved in the electrochemical tests in Figure 2, whose resulting capacities are displayed in Figure 3a. Minor discrepancies observed can be attributed to non-faradaic currents associated with parasitic reactions, as indicated in Figure 2. The difference in the amounts of Li stored during weak and strong lithiation regimes does not directly translate into an equivalent gap between capacity densities (compare the gaps between the bars in Figure 4e with those in Figure 4f). This indicates that lithiation occurs through two processes: (i) the increase in Li concentration within the lithiated regions up to saturation, and then (ii) the propagation of the lithiation front into the component volume. For instance, while the amount of Li stored in Ag increases significantly from (weak) $\approx 4\times 10^{17}$ to (strong) $\approx 2\times 10^{19}$ atoms, there is no evidence of a change in capacity density within the quoted uncertainties (from 3.05 ± 0.59 to 2.50 ± 0.03, both $\times 10^{22}$ Li at. $cm^{-3}$) as the lithiation range extends from $\approx 300$ nm (see the down arrow in Figure 4c) to over 4.5 µm (Figure 4d).

Figure 4f shows the capacity densities of metal components calculated from the lithiation ranges of the Li depth profiles in Figure 4c and 4d. This calculation provides more reliable capacity values than those inferred from electrochemical tests, considering the full component volume. To illustrate this effect, Figure 4f (right axis) shows the probable range of specific capacities for each metal component. This range, represented as line segments, includes specific capacities between the lowest capacities by electrochemistry and the highest ones by NRA. For Li depth profiles exceeding the penetration limit for 3 MeV H$^+$ (Sn, Al and Ag), NRA tends to overestimate the specific capacity by excluding the dilute concentration regions that result from the propagation of the lithiation front, which is considered the rate-limiting mechanism. Metals where lithiation involves surface accumulation of Li by SEI or plating (such as weakly lithiated Zn and Ag, and Cu, as suggested by electrochemical tests and NRA results) exhibit capacity densities close to the atomic density of metal Li ($\sim 4.6\times 10^{22}$ Li at. $cm^{-3}$) and indefinite specific capacities (marked with "*" in Figure 4f) for Li absorption.

Ion beam analysis with protons is an effective technique for direct detection of Li by NRA.[50,51] However, its depth resolution for detecting other elements via RBS from proton backscattering is limited by the weak interaction of protons with the metal component lattice. This results in small stopping powers and non-Rutherford collision sections with some studied metals, (such as Mg and Al). In contrast, RBS using He$^+$ offers complementary advantages and disadvantages. In this study, Li is also detected indirectly through He$^+$ RBS via metal dilution (**Figure 5**). This method has a higher error margin due to the detection accuracy of potential contaminants (e.g., oxygen and carbon from organic solvents in the electrolyte) but offers an order-of-magnitude improvement in depth resolution (60 nm by He+ RBS vs. 800 nm by H+ NRA, as estimated by RESOLNRA 1.11 code). Consequently, we have decided to exclude H+ RBS measurements (which are acquired simultaneously with NRA measurements) from our analysis.

Figure 5a displays the RBS spectra for pristine, weakly lithiated, and strongly lithiated metal components. The depth dependence of the inelastic losses of the kinetic energy of backscattered He+ is also modeled using the SIMNRA 7.03 code. This modeling allows the conversion of the energy axes in Figure 5a to depth within the metal components, as shown in Figure 5b and 5c. Depth is measured from the surface of the metal components without considering the Li plating on top. The relative position of this inner surface to the surface of the sample is estimated from the analysis of the Li depth profiles by RBS in the SI.

In general, the RBS spectra of the lithiated metal components exhibit lower intensities than those of the pristine ones. In the low lithiation regime, this trend aligns with the formation of substitutional solid solutions, where host metal atoms are partially replaced, effectively reducing the host metal signal. This is in contrast to interstitial incorporation, where the metal intensity would be expected to remain unchanged. The decrease in RBS intensity in the high lithiation regime is more complex to address, as it can be attributed to multiple factors. A high insertion of Li, even in interstitial positions, can trigger the formation of alloys and low-density subdomains with Li-Li lattices (e.g., in lithiated Mg, as discussed below). The NRA and RBS spectra shown in Figures 4a, 4b, and 5a are consistent with each other. The lithiation behaviors interpreted from the composition profiles achieved by combining the data from these two techniques agree with the information inferred from the electrochemical tests in Figure 2.

Figure 5b and 5c illustrate color gradient representations of the $[Li]_{1-y}[M]_y$ composition depth profiles, where [ ] denotes the atomic concentration, for both weakly and strongly lithiated metal components. Surface regions have been zoomed in to identify the processes operating during electrochemical cycles leading to surface Li accumulation. Although we cannot distinguish directly from IBA results between the SEI, Li plating, and dead Li (the irreversible fraction of the plating) due to the small differences in their densities, we can predict the chemical environment of the adsorbed Li using other arguments. Since SEI formation is a self-limiting reaction that, once fully formed, protects the component surface from further reactions, it presumably operates during the first electrochemical cycles, extending to subsequent cycles under weak lithiation conditions. Consequently, we can assume that the small surface accumulation of Li during weak lithiation (Figure 5b: subsurface Li forms films up to 100-nm thick on Zn, Ag, and Cu) consists mainly (but not exclusively) of Li trapped in the SEI, while the large accumulation during strong lithiation (Figure 5c: 1.5 μm-thick Li plating on Cu) corresponds mainly to plated Li. Metals forming pure alloys with Li (Sn and Al) show neither plating nor SEI formation. The minimal plating observed on Mg in the strong lithiation regime suggests a kinetic rather than thermodynamic origin. As mentioned above, the low-energy tails of Li depth profiles by He$^+$ RBS, which offer better depth resolution, are analyzed in the SI.

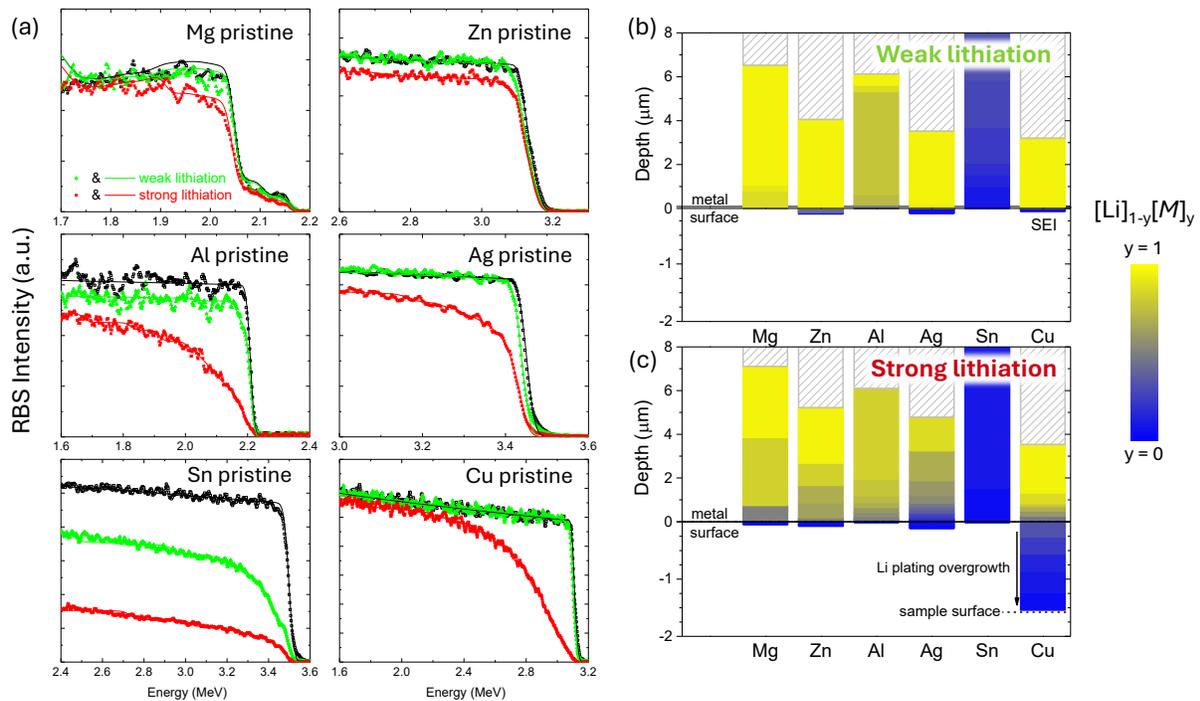

**Fig. 5 (a)** RBS spectra for pristine (black), weakly (green) and strongly (red) lithiated metal components, with symbols indicating experimental data and curves representing fits obtained using SIMNRA 7.03 code. The difference between the pristine and lithiated RBS spectra provides an estimate of metal dilution due to Li intercalation into both the metal component surface and lattice (details in the SI). **(b)** and **(c)** Composition depth profiles for weakly and strongly lithiated metal components. The zebra-patterned segments indicate the penetration limits for 4-MeV He$^+$. Depth scales have been enlarged in the surface region (negative values) to illustrate the effects of SEI formation and Li plating.

**Figure 6** shows the Li content, as imaged by scanning electron microscopy (SEM) using secondary (SE) and backscattered electrons (BSE), at different depths in strongly lithiated metal components. BSE-SEM images provide a comparison between the surface (outside the eroded area) and the bulk of the lithiated metal component (inside) in terms of BSE intensity contrast. Since the BSE signal increases with the atomic number ($Z$) of the target atoms, Li absorption reduces the BSE intensity. Consequently, signals from Li-rich regions appear less intense (bluer) compared to those from Li-poor regions (yellower). BSE intensity contrast should be assessed qualitatively rather than quantitatively due to the following reasons: (a) BSE signal includes a morphological contribution, causing the BSE intensity to decrease with increasing surface roughness of the eroded areas. (b) The implantation range of Ga+ ions in the metal component bulk depends on Li content. Li, being a light metal, has a lower sputtering yield and a larger implantation range than other metals. Thus, implanted Ga increases the BSE signal in pure metals, where the implantation is shallower compared to lithiated metals. Although we qualitatively understand that roughness and Ga implantation affect the BSE intensity (decreasing and increasing it, respectively), quantifying these effects is challenging. (c) The erosion rate by FIB milling varies with the Li content, which in turn changes with the component thickness. This results in significant uncertainty in the depth of erosion. Details regarding Ga+ interactions with the metal components, including sputtering yield and implantation range, are provided in the SI. It is crucial to explain why standard SEM-BSE analyses of solid solutions and/or alloys in cross-section are often unreliable. In this direction, the Li content varies continuously, leading to small incremental changes in BSE signal contrast. These changes are typically smaller than the variations in BSE signal caused by wavy roughness induced by FIB milling on the walls of the eroded area (more details in the SI).

The uncertainties described above force us to group the distinctive behaviors of Al, Ag and Cu (shown in Figure 6) into two Ga$^+$ dose ranges: low (2-4 x$10^5$ μC cm$^{-2}$) and high (1-2 x$10^6$ μC cm$^{-2}$) dose. Metals forming pure alloys with Li (here Al) do not exhibit a significant change in BSE intensity contrast with Ga$^+$ dose, indicating a roughly uniform Li depth profile. For Ag, where Li is initially adsorbed and then absorbed/alloyed, low doses do not produce significant in BSE intensity, while high doses are sufficient to partially uncover the buried surface of the pure metal. In metals where Li is purely adsorbed (e.g., Cu), low doses are enough to uncover the surface of the pure metal. These results, although qualitative, agree with those obtained by NRA and RBS.

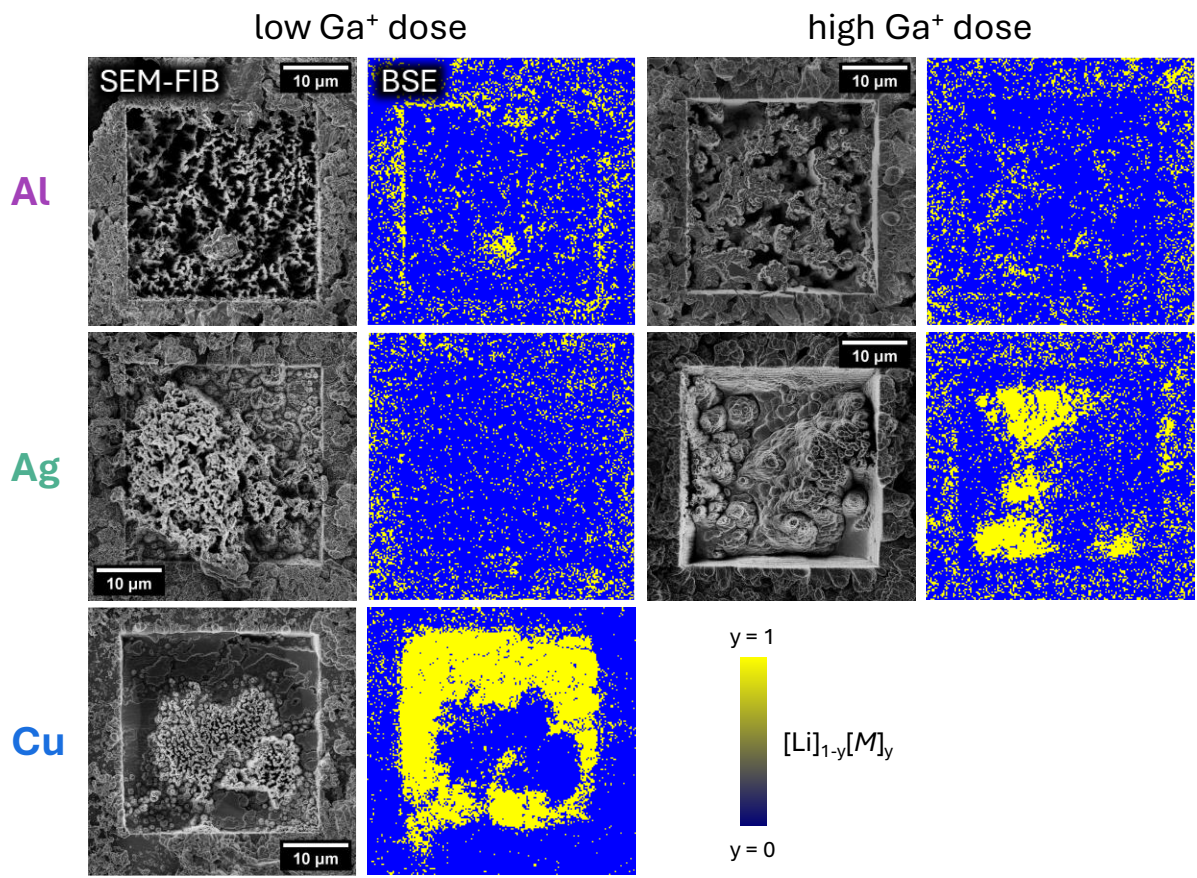

**Fig. 6** SE- and equalized BSE-SEM images of strongly lithiated metal components of Al, Ag and Cu, whose volume has been uncovered by erosion using 30 keV Ga⁺ FIB milling with doses ranging from 2-4x10⁵ (low dose) to 1-2x10⁶ µC cm⁻² (high dose). The BSE intensity contrast is interpreted in terms of Li content from blue (Li-rich regions) to yellow (Li-poor metal regions).

The results of this study have been analyzed in conjunction with the binary Li-*M* phase diagrams (summarized in **Figure 7**) under normal temperature and pressure conditions (NTP, i.e., 25 °C at 1 atm). These NTP conditions correspond to those in which the investigated metal components were electrochemically lithiated/delithiated. Metals forming pure alloys with Li (Al, Sn and Zn, the latter exhibiting slow lithiation kinetics) show Li miscibility lower than 3% before Li-*M* alloys start forming. A further increase in Li content leads to transitions between alloys with higher Li proportions, which implies a greater Li storage capacity, until metal Li begins to segregate from the overlithiated metals when the atomic Li content reaches 50% for Zn, 69% for Al and 82% for Sn. Since lithiation proceeds from the electrochemical reduction of Li ions flowing through the electrolyte, rather than being segregated, excess Li is not incorporated into the bulk component, resulting in plating (indicated by the right arrows). Although the Li-Zn, Li-Al, and Li-Sn phase diagrams predict Zn lithiation thermodynamics similar to those of Al and Sn, they do not address the differences in their lithiation kinetics (as observed in Figure 4b by comparing the intensities of Li NRA bands for Zn, Al, and Sn). A plausible explanation for the slow lithiation kinetics of Zn, consistent with the findings of Wang *et al.*[49], could be the premature formation of a single alloy phase of $LiZn_4$ (with a Li proportion of 20%) at 15 at.% of Li (marked with "*"). This early formation consumes Li by filling vacancies, hindering its bulk diffusion. This hypothesis is supported by the behavior of the CV curves during the weak lithiation of Zn (Figure 2), where several peaks attributed to different alloys are identified, experiencing small changes in proportions relative to each other.

In the case of metals creating solid solutions with Li (Ag and Mg, as revealed in Figure 4b from the low-energy tails of the Li NRA bands), the miscibility range of Li in the host metal lattice (marked with two-arrow segment in Figure 7) extends up to 47 at.% in Ag, which is exceptionally wide, and 18 at.% in Mg under NTP conditions. A further increase in Li content in Ag leads to alloy formations, while the segregation of metal Li from overlithiated Ag is minimal and occurs only when Li at.% exceeds 93%. On the other hand, the increase in Li concentration in Mg beyond 18 at.% results in a phase transformation from the metal Mg lattice (S.G. *P6₃/mmc*) to the metal Li lattice (S.G. *Im-3m*),[52] which is completed when the Li content exceeds 33 at.%. This means that the solid solution of Li in the Mg lattice changes to Mg in the Li lattice, as Mg and Li are fully miscible and do not form an alloy. This early transformation of the component lattice (marked with "*") may explain the slow lithiation kinetics of Mg, as well as its moderate Coulombic efficiency (< 90% in Figure 3c).

Cu-Li phase diagram (bottom row in Figure 7) reveals that Cu neither forms alloys nor massive solid solutions with Li. The Cu lattice can accommodate up to 15 at.% of Li under NTP conditions, while excess Li remains on the surface during Li electroplating.

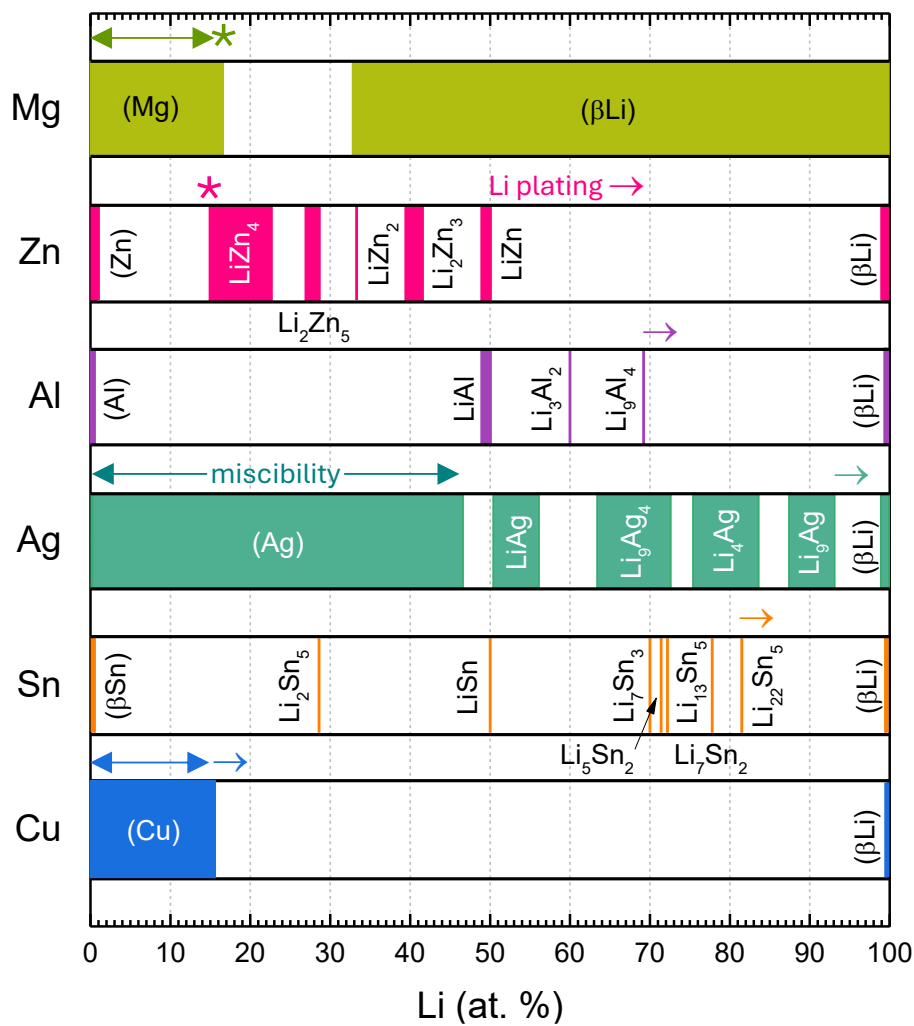

**Fig. 7** Binary phase diagrams for the systems (from top): Li-Mg, Li-Zn, Al-Li, Ag-Li, Li-Sn and Cu-Li, based on references[53–58], under NTP conditions. The colored regions correspond to individual phases whose composition is labelled, while the white regions indicate phase mixtures. The marks above the diagrams denote: two-arrow segment (miscibility range), right arrow over white region (start of Li plating), and "*" in the Li-Mg and Li-Zn diagrams (limitations to the lithiation kinetics discussed in the text).

At this point, we use Density Functional Theory (DFT) to perform ab—initio calculations to address the lithiation behaviors detected in the studied metal components through a combination of electrochemical tests (CV and GCD) and complementary ion-beam-based techniques (NRA and RBS). The calculations also aim to determine the adsorption ($E_{ads}$) and absorption energies ($E_{abs}$) of Li in the metal lattice, in order to compare them with the electrochemical potentials of lithiation and delithiation referenced in Figure 2. For this comparison, we propose an energy landscape model for the electrochemical lithiation of metal components, as shown in **Figure 8**.

According to the energy landscape model in Figure 8a, $E_{ads}$ (written in blue) refers to the energy difference between the Li-ion states within the electrolyte $E_{ele}^{Li^+}$ and the energy of Li adsorbed on the metal surface $E_{int}^{Li}$ after it is electrochemically reduced. Consequently, $E_{ads}$ can be correlated with the lithiation voltage, $E_{ads} = -eV_l$, for a dilute surface concentration of Li.[59] The term $E_{int}^{Li}$ accounts for the replacement of the surface metal atom with Li ($E_{surf}^{Li} - E_{surf}^{M}$) as well as the formation of the Li/metal interface ($E_{int}^{Li/M}$), so that for a Li-on-Li scenario $E_{int}^{Li} = 0$. On the other hand, $E_{abs}$ (written in red) is defined as the difference between $E_{ele}^{Li^+}$ and $E_{M}^{Li}$, the energy associated with Li absorption within the metal bulk. In the Li-on-Li case, $\Delta E_{sur} = E_{ads} - E_{abs}$ estimates the excess energy associated with the surface of a solid of Li (here $E_{surf}^{Li} \approx -1.55$ eV at.⁻¹ from our ab—initio calculation). $E_{M}^{Li}$ encompasses two types of Li absorption states: (1) the solid solution, denoted as $E_{ss}^{Li_x M_{1-x}}$, where Li gradually dissolves into the host metal lattice, and (2) the alloy state, $E_{a}^{Li-M}$, wherein Li and the metal form various compounds with distinct lattice structures. The solid solution includes an energy strain contribution $\Delta E_{ss}^{Li_x M_{1-x}} \big]_{x \to 0}^{x}$, which increases as the Li content rises due to differences in the Goldschmidt atomic diameters of Li and the metal. The limit of solubility of Li in the host metal lattice corresponds to the Li content $x_s$ that induces a strain energy that exceeds $\Delta E_{sur}$ and/or the alloy formation barrier $\phi_a$. Excess Li, which would give rise to forbidden solutions with $x > x_s$, remains adsorbed on the metal surface (i.e. this Li and the metal become immiscible as in the case of Cu) or triggers the formation of *Li-M* alloys through the restructuring of the compound lattice to minimize strain energy (e.g. for Li contents >3 at. % in Al and >47 at. % in Ag, as shown in Figure 7). Further solid solutions in alloy lattices appear when the Li content exceeds the Li stoichiometry of the alloy compound, giving rise to transitions towards alloys with higher Li/*M* ratios (e.g. LiAl → Li₃Al₂ → Li₉Al₄ in Figure 7) until metal Li is segregated.

Figure 8b displays the values of $E_{ads}$ and $E_{abs}$ derived from ab—initio calculations for diluted Li concentrations in the metal components. The adsorption and absorption sites of Li within the metal lattices used for the calculation are detailed in the SI, along with a description of the approximations and tools used. Note that the ab—initio calculations qualitatively predict that all metal components investigated, except Cu, show a clear tendency to form solid solutions or alloys with Li (with $E_{ads} > E_{abs}$). On the other hand, the quantitative comparison between computed $E_{ads}$ and $E_{abs}$ and electrochemical voltages referenced in Figure 2 requires considering that, unlike the ab—initio calculations, the electrochemical tests incorporate lithiation kinetics. Thus, for example, slow Li diffusion into the metal bulk (e.g. in Ag and Zn) with the consequent surface accumulation of Li makes the electrochemical estimation of the lithiation voltage $V_l$ compared with $E_{ads}/e$ unfeasible. The electrochemical voltages affected by the process kinetics are enclosed in parentheses in Figure 8b. Another factor to consider is that the ab—initio calculations have been developed for dilute Li concentrations, such that the adsorbed (or absorbed) Li atoms do not interact with each other. This limits the range of lithiation capacities in which the comparison makes sense, particularly between $eV_l$ and $E_{ads}$. It also does not allow consideration of the structural changes involved in alloy formation.

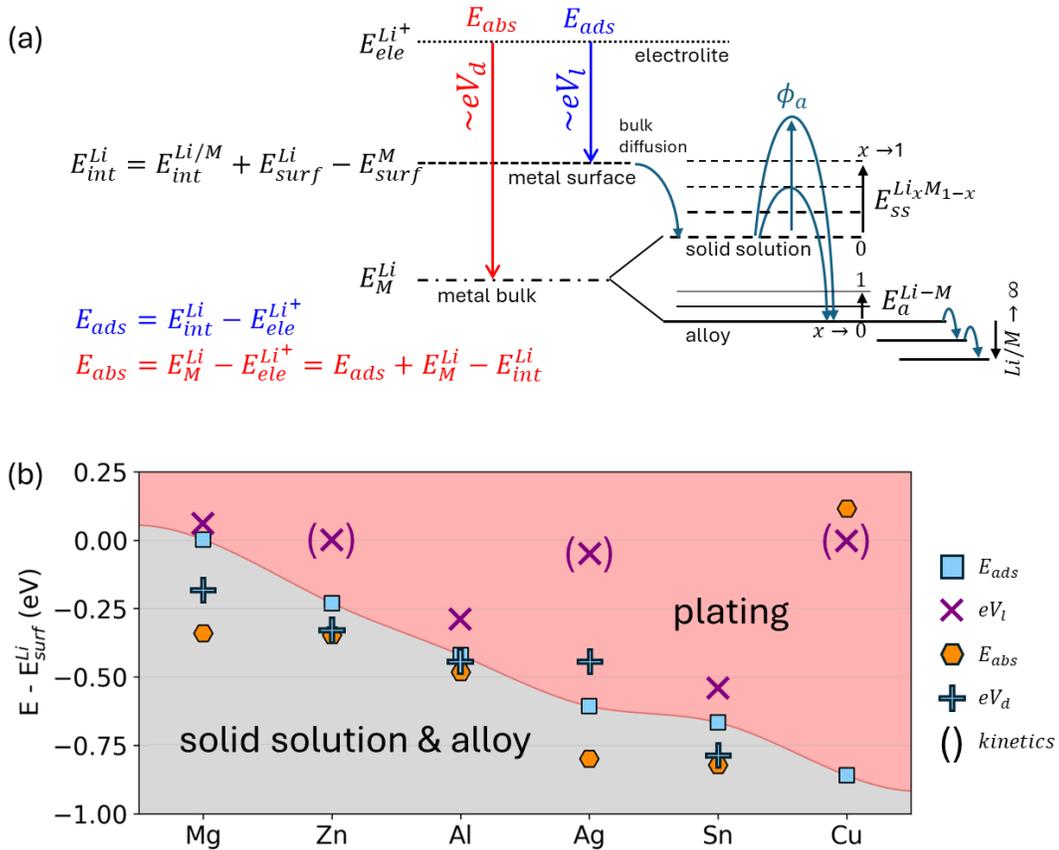

**Fig. 8** Ab—initio calculation of the lithiation thermodynamics for metal components. **(a)** An energy landscape model proposed to interpret the energies of adsorption $E_{ads}$ and absorption $E_{abs}$ of Li, illustrated by transitions between states using blue and red arrows, in terms of lithiation $V_l$ and delithiation $V_d$ voltages, respectively. **(b)** The ab—initio computed values of $E_{ads}$ and $E_{abs}$, relative to the surface energy of a solid of Li, for diluted Li in different metal components. These values (in eV) are compared with the electrochemical voltages (arealess symbols) referenced in Figure 2, with "()" indicating those influenced by the lithiation kinetics. The curve connecting $E_{ads}$ values delineates the distinction between different lithiation behaviors: plating for $E_{ads} < E_{abs}$, and the formation of solid solutions and alloys otherwise.

Comparison of experimental and computational data ($eV_l$ and $eV_d$ with $E_{ads}$ and $E_{abs}$, respectively) in Figure 8b leads to the following conclusions: (i) Metal forming pure alloys with Li (Sn, Al and Zn) show good agreement between $E_{abs}$ and $eV_d$, while the agreement between $E_{ads}$ and $eV_l$ depends on the lithiation kinetics (i.e., faster kinetics, e.g. for Sn and Al, implies better agreement). (ii) Metal creating solid solutions with Li (Mg and Ag below the limit of solubility) show poorer concordance with $eV_d > E_{abs}$. This discrepancy may arise because the ab—initio calculations do not account for repulsive interactions between Li absorbates, making the computed Li absorption appear more favorable than observed experimentally. The fact that the Mg lattice transforms into the Li lattice for a Li content as low as 33 at. % explains why $V_l$ drops to negative values in the GCD curves (Figure 2). (iii) Metals where Li remains adsorbed (Cu), giving rise to Li-on-Li growth, do not provide experimental results on Li adsorption and absorption.

Finally, the energy landscape model in Figure 8a establishes that, for metals forming alloys with Li, $E_{ads} - E_{abs} > \phi_a$ (the alloy formation barrier). Consequently, as the gap between $E_{ads}$ and $E_{abs}$ (and hence $\phi_a$ as well) is smaller, it becomes easier to reach the solubility limit that triggers alloy formation. According to these predictions, the metals that form pure alloys with Li (Al, Zn and Sn) have the lowest $E_{ads} - E_{abs}$ values; whereas, for metals with a larger gap between $E_{ads}$ and $E_{abs}$ (Mg and Ag), the solid solutions containing Li persist for longer.

## 3. Conclusion

Our findings, which integrate electrochemical tests with ion beam analysis, reveal three distinct thermodynamic lithiation behaviors in metal components (summarized in Table I). These behaviors are consistent with the binary Li-*M* phase diagrams of the studied systems, with existing reports,[1,2] and their electrochemical features fit well to fundamental thermodynamic parameters derived from our ab—initio simulations. By leveraging these references and the similarities in the binary phase diagrams, we can extend our results to a wider range of metal components[49,52,60–65] indicated with an asterisk in Table I.

Beyond the thermodynamics provided by the phase diagrams in Figure 7, our experiments demonstrate that differentiating between lithiation behaviors requires investigating the lithiation kinetics of each system, as it governs the initial stages of Li adsorption and absorption. Our results indicate that such kinetics are not closely related to the lithiation habit. For example, while Zn—which thermodynamically tends to alloy—and Mg—which forms solid solutions with Li—exhibit qualitatively similar behavior in the weak lithiation regime (as does Ag with Cu, see Figure 3 and 5c), they diverge significantly in the strong lithiation regime (Figure 4d). The limitations to the lithiation kinetics in these systems (Zn and Mg) have been tentatively linked to early transformations of the lattice component into single solid phases (specifically, $LiZn_4$ alloy with 5 at.% Li vacancies, and solid solutions containing 66 at.% Mg in the Li lattice) at Li contents as low as 15 at.%. Further studies addressing the kinetics and degree of reversibility of the lithiation mechanisms in metal components are currently under progress.

Finally, lithiation depth profiling using ion-beam based techniques provides comprehensive data with greater accuracy than those obtained using electrochemistry, such as the maximum capacity of a metal component to store Li in terms of its volume utilization. This becomes feasible once the dependence of the Li depth profile on lithiation conditions is understood. The results reported here represent a step forward in the control and engineering of metal components as multifunctional electrodes for the implementation of lithium-ion batteries with simple, optimized designs and improved energy densities.

| Lithiation behavior | Metal | Details |
|---|---|---|
| Metals forming pure alloys with Li | Al<br>Sn<br>Zn<br>In*<br>Pb*<br>Au* | **Pros:**<br>  o  higher Li storage capacity<br>  o  negligible Li losses by SEI formation<br>  o  no dendritic overgrowth<br>  o  high oxidation voltage ($\sim\frac{1}{2}[V_l + V_d]$)—suitable as anode<br><br>**Cons:**<br>  o  low lithiation/delithiation reversibility (by coulombic efficiency) during alloy formation<br>  o  brittleness and propensity to mechanical fatigue (details in the SI), with capacity fading |
| Metals forming solid solutions with Li | Ag<br>Mg<br>Pt*<br>Cd* | **Pros:**<br>  o  better lithiation/delithiation reversibility once the SEI is formed<br>  o  thin Li plating with moderate dendritic overgrowth<br>  o  the components retain their toughness under strong lithiation regime<br><br>**Cons:**<br>  o  lower Li storage capacity<br>  o  Li losses by SEI formation<br>  o  oxidation voltage highly dependent on lithiation kinetics |
| Li immiscible metals | Cu<br>stainless steel* | **Pros:**<br>  o  better lithiation/delithiation reversibility once the SEI is formed<br>  o  inert to Li redox—suitable as a non-reactive component (e.g. current collector acting as Li diffusion barriers).<br><br>**Cons:**<br>  o  thick Li plating with enhanced dendritic overgrowth<br>  o  mechanical failures of loss of contact in solid-state LIBs |

**TABLE I.** Lithiation behaviors in metal components. *) lithiation habits taken from the literature.[49,52,60–65]

## 4. Experimental Section / Methods

**Figure 9** illustrates the flow diagram of the experiments carried out, which include: (a) lithiation of the metal foils through electrochemical tests, (b) extraction for post-mortem characterizations, and (c) analysis via interaction with ion beams.

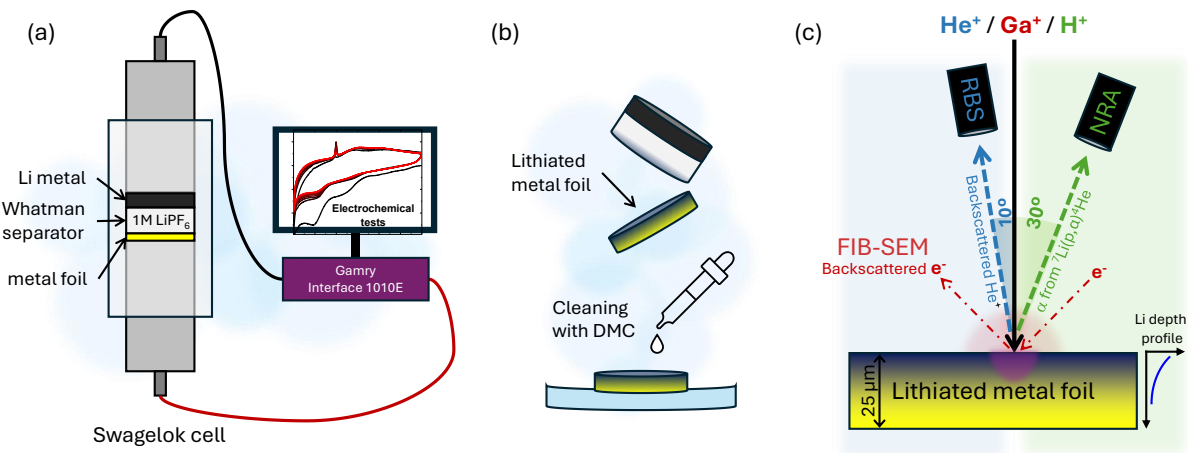

**Fig. 9**: Flow diagram of the studies encompassing the following steps: (a) sample preparation through electrochemical cycling, (b) sample extraction, and (c) post-mortem analysis using ion beams.

    All metal components were used as 25 μm-thick foils with a purity greater than 98%, supplied by Goodfellow. The components were cut with a high-precision cutter (EL-Cut by EL-CELL) for 12-mm diameter circular electrodes. The electrochemical cells consisted of metal components separated from the Li source (in this case, 99.9%-pure metal Li foil by S4R) by a Whatman GF/B glass microfiber filter (Sigma Aldrich), which was moistened with 20 μl of a 1.0 M LiPF$_6$ electrolyte in a solvent mixture of ethylene carbonate, diethyl carbonate and dimethyl carbonate (EC/DEC/DMC with a volume ratio 1:1:1, by Sigma Aldrich). All components were encapsulated within Swagelok-type two-terminal cells (S4R). Their electrochemical performances were investigated after a period of 12 hours from assembly, monitoring the $V_{oc}$ to ensure its stabilization. Given that lithiation behavior varies with Li concentration, we investigated two different regimes: weak lithiation and strong lithiation. Weak lithiation was performed using cyclic voltammetry (CV) within an electrochemical window of 0-3 V (0-2.5 V for Zn and Sn, as they exhibited anomalous reactions for voltage values close to 3 V), in which the open circuit voltages ($V_{oc}$) of all studied cells are included, at a scan rate of 0.5 mV s$^{-1}$. In contrast, strong lithiation was achieved through galvanostatic charge-discharge (GCD) at a constant current of 400 μA up to a maximum charge of 4 mAh. Electrochemical tests (CV and GCD) were conducted using Gamry

Interface 1010e potentiostats. After lithiation, metal components were extracted from the cells and rinsed with battery-grade DMC (Sigma-Aldrich) for post-mortem studies. The sample preparation, extraction and mounting in the corresponding sample holders (adapted for each experiment) were performed in an Ar-filled Jacomex glovebox. The samples were transported to the analysis setups in sealed aluminum bags.

Both NRA and RBS measurements were conducted using a tandetron accelerator operating at a maximum terminal voltage of 5 MV at CMAM (UAM, Madrid, Spain)[66] utilizing different projectiles: 3-MeV H+ for NRA and 4-MeV He+ for RBS. The ion beams were extracted from the accelerator with a probe size ranging from 1—1.5 mm$^2$ (divergence of ~0.1°) and collimated by a pair of slits, whose openings were modified according to the deadtime in the detectors, conditioned by the nature of the sample. The ion beam currents were calibrated in a Faraday cup and then measured in the sample holder during the experiments. Currents of 6-8 nA were used for NRA (6-12 nA for RBS) until cumulative charges of 6 μC (3 μC) were reached. Two Si barrier particle detectors (by ORTEC) were positioned so that the detector used for He+ RBS (12 keV energy resolution) was at a scattering angle of 170° and the detector for H+ NRA (18 keV energy resolution) at 150°, both with respect to the direction of the incident ion beam, which was normal to the sample with a tilt of 2° to avoid channeling. NRA and RBS spectra are fitted using the SIMNRA 7.03 code,[46,47] with collisional cross-sections reported by Paneta *et al.*[67], while their depth resolutions were estimated for our samples and measurement conditions using RESOLNRA 1.11 code.[48]

The in-depth analysis by FIB-SEM was carried out on areas of 30x30 μm$^2$, which were eroded with a 30 keV Ga+ focused ion beam (in a FIB-SEM ionLine system by Raith GmbH) with doses ranging from 0.2—2 x10$^6$ μC cm$^{-2}$. FIB milling and SEM characterization alternated in-situ. Ion-beam analyses were performed on discharged (i.e., lithiated) metal components, where lithiation comprises both reversible and irreversible cumulative contributions.

# Supporting Information

**Supporting Information available:**

- Comprehensive overview of NRA and RBS spectra processing and correlation methodology to detect Li depth profiles and analyze their low-energy tails (i.e., the deepest Li layers)
- Details on the interaction of the lithiated metal components with Ga+ beam used in FIB-SEM, focusing on sputtering yield and implantation depth, as well as how FIB-induced morphology contributes to the BSE signal
- Information on ab-initio simulations
- Insights into the brittle mechanical properties of lithiated metal components

# Acknowledgements

This work was supported by the European Union through the Horizon Europe Research and Innovation Program under grant agreement No. 101103834 (Project OPERA). It also received funding from the Spanish national Agencia Estatal de Investigación as part of the following programs: M-ERA.NET Call 2021 (Project SOLIMEC, subprojects PCI2022-132955 and PCI2022-132998), "Proyectos de Generación de Conocimiento" (Project NanoCatCom, Ref. PID2021–1246670B) and "María de Maeztu" Programme for Units of Excellence in R&D (CEX2023-001316-M). We acknowledge the service from the MiNa Laboratory at IMN, and funding from CM (project S2018/NMT-4291 TEC2SPACE), MINECO (project CSIC13-4E-1794) and EU (FEDER, FSE). We are grateful to the UK Materials and Molecular Modelling Hub for computational resources, which is partially funded by EPSRC (EP/P020194/1, EP/W032260/1). We also acknowledge the support from the Center for Micro-Analysis of Materials (CMAM) — Universidad Autónoma de Madrid, for the beam time proposals, with codes STD011/24 and STD032/24, and its technical staff for their contribution to the operation of the accelerator. As well as the support from ReMade@ARI support "Funded by the European Union as part of the Horizon Europe call HORIZON-INFRA-2021-SERV-01 under grant agreement number 101058414 and co-funded by UK Research and Innovation (UKRI) under the UK government's Horizon Europe funding guarantee (grant number 10039728) and by the Swiss State Secretariat for Education, Research and Innovation (SERI) under contract number 22.00187. Views and opinions expressed are however those of the author(s) only and do not necessarily reflect those of the European Union or the UK Science and Technology Facilities Council or the Swiss State Secretariat for Education, Research and Innovation (SERI). Neither the European Union nor the granting authorities can be held responsible for them."